\documentclass[final]{article}

\PassOptionsToPackage{numbers}{natbib}

\usepackage[final]{neurips_2024}

\usepackage[utf8]{inputenc} %
\usepackage[T1]{fontenc}    %
\usepackage{hyperref}       %
\usepackage{url}            %
\usepackage{booktabs}       %
\usepackage{amsfonts}       %
\usepackage{nicefrac}       %
\usepackage{microtype}      %
\usepackage{xcolor}         %
\usepackage{graphicx}
\usepackage{soul}
\usepackage{CJK}

\title{Beyond Causal Discovery for Astronomy: Learning Meaningful Representations with Independent Component Analysis}

\author
{Zehao Jin (金泽灏)$^{1,2,3,\ast}$,
Mario Pasquato$^{3,4,5,6,7,\dagger}$,
Benjamin L.\ Davis$^{1,2,\dagger}$,
\\
\textbf{Andrea Valerio Macci\`{o}}$^{1,2,8}$,
\textbf{Yashar Hezaveh}$^{3,4,5,9}$\\
\\
\normalsize{$^{1}$New York University Abu Dhabi,}\\
\normalsize{P.O.\ Box 129188, Abu Dhabi, United Arab Emirates.}\\
\normalsize{$^{2}$Center for Astrophysics and Space Science (CASS), New York University Abu Dhabi,}\\
\normalsize{P.O.\ Box 129188, Abu Dhabi, United Arab Emirates.}\\
\normalsize{$^{3}$Ciela Institute, Montr\'{e}al, Canada.}\\
\normalsize{$^{4}$Mila - Quebec Artificial Intelligence Institute,}\\
\normalsize{6666 Rue Saint-Urbain, Montr\'{e}al, Canada.}\\
\normalsize{$^{5}$D\'{e}partement de Physique, Universit\'{e} de Montr\'{e}al,}\\
\normalsize{1375 Avenue Th\'{e}r\`{e}se-Lavoie-Roux, Montr\'{e}al, Canada.}\\
\normalsize{$^{6}$Dipartimento di Fisica e Astronomia, Universit\`{a} di Padova,}\\
\normalsize{Vicolo dell'Osservatorio 5, Padova, Italy.}\\
\normalsize{$^{7}$Istituto di Astrofisica Spaziale e Fisica Cosmica (INAF IASF-MI),}\\
\normalsize{Via Alfonso Corti 12, I-20133, Milan, Italy.}\\
\normalsize{$^{8}$Max-Planck-Institut f\"{u}r Astronomie, K\"{o}nigstuhl 17, Heidelberg, 69117, Germany.}\\
\normalsize{$^{9}$Center for Computational Astrophysics, Flatiron Institute,}\\
\normalsize{New York, NY, United States of America.}\\
\\
\normalsize{$^\ast$Corresponding author. Email: \href{mailto:zj448@nyu.edu}{zj448@nyu.edu}}\\
\normalsize{$^\dagger$These authors contributed equally to this work.}
}

\begin{document}
\begin{CJK*}{UTF8}{gbsn}

\maketitle
\end{CJK*}

\begin{abstract}
We present the first steps toward applying causal representation learning to astronomy.
Following up on previous work that introduced causal discovery to the field for the first time, here we solve a long standing conundrum by identifying the direction of the causal relation between supermassive black hole (SMBH) mass and their host galaxy properties.
This leverages a score-based causal discovery approach with an exact posterior calculation.
Causal relations between SMBHs and their host galaxies are further clarified by Independent Component Analysis (ICA).
The astrophysical problem we focus on is one of the most important open issues in the field and one that has not seen a definitive resolution in decades.
We consider the space of six physical properties of galaxies, subdivided by morphology: elliptical, lenticular, and spiral, plus SMBH mass.
We calculate an exact posterior over the space of directed acyclic graphs for these variables based on a flat prior and the Bayesian Gaussian equivalent score. 
The nature of the causal relation between galaxy properties and SMBH mass is found to vary smoothly with morphology, with galaxy properties determining SMBH mass in ellipticals and vice versa in spirals.
This settles a long-standing debate and is compatible with our theoretical understanding of galaxy evolution.
ICA reveals a decreasing number of meaningful Independent Components (ICs) from ellipticals and lenticular to spiral.
Moreover, we find that only one IC correlates with SMBH mass in spirals while multiple ones do in ellipticals, further confirming our finding that SMBH mass causes galaxy properties in spirals, but the reverse holds in ellipticals.

\end{abstract}

\section{Introduction}
Astronomy is an observational science and, as such, relies on empirical correlations to test theories.
This gives rise to conundrums about the causal significance of measured quantities that cannot be resolved through experimental intervention.
An example is the long-standing debate about the causal interpretation of the correlations between the mass of central BHs and the properties of their host galaxies \citep[][]{Magorrian:1998,Silk:1998,Ferrarese:2000,Gebhardt:2000,DiMatteo:2005,Sijacki:2007,DiMatteo:2008,Schaye:2010,Kormendy:2013,Gaspari:2013,Heckman:2014,Soliman:2023}.
Recently, causal discovery techniques have been proposed as a way out of this impasse \citep[][]{Pasquato:2023, pirsa_PIRSA:24020094, JinSubmitted} and, more generally, there is increasing interest in treating astronomical problems using causality methods \citep[see e.g.,][]{2019RNAAS...3..179P, 2021ApJ...923...20P}.

However, this growing body of work typically takes for granted that the variables measured by astronomers are suitable for uncovering causal relations, without engaging with the problem of finding better representations for astronomical data in the spirit of causal representation learning.
Our contribution is focused on filling this gap, albeit in a somewhat limited fashion, by applying ICA to the space of galaxy properties introduced by \cite{Pasquato:2023}.
We show that the resulting ICs have a causal relation with the mass of the SMBH hosted by each galaxy, which depends on galaxy morphology.
Namely, in elliptical galaxies the mass of the SMBH is an effect of multiple ICs representing combinations of observed galaxy properties, while in spiral galaxies the mass of the SMBH causes a single IC, which likely represents a coordinate corresponding to the physical coupling mechanism between black hole feedback and galaxy growth.
This is in agreement with theoretical expectations from galaxy formation simulations \cite[e.g.,][]{Springel:2005} and settles the debate on the causal interpretation of the correlations between the mass of central SMBHs and the properties of their host galaxies.
Moreover, our contribution constitutes the first advancement toward applying causal representation learning to astronomy.

\section{Sample}\label{sec:sample}

To explore the causal relationship between SMBHs and their host galaxies, we use the state-of-the-art dataset of a sample of 101 galaxies and their dynamically-measured SMBH masses.
The dataset comprises seven variables of interest: dynamically-measured black hole mass ($M_\bullet$), central stellar velocity dispersion ($\sigma_0$), effective (half-light) radius of the spheroid\footnote{
Throughout this article, we use the terms ``bulge'' and ``spheroid'' interchangeably to refer to the spheroid component of spiral and lenticular galaxies or the entirety of pure elliptical galaxies.
} ($R_\mathrm{e}$), the average projected density within $R_\mathrm{e}$ ($\langle{\Sigma_\mathrm{e}}\rangle$), total stellar mass ($M^*$), color ($W2-W3$), and specific star formation rate (sSFR). 
Among these seven variables, $\sigma_0$, $R_\mathrm{e}$, and $\langle{\Sigma_\mathrm{e}}\rangle$ cover the fundamental plane of elliptical galaxies \cite{Djorgovski:1987}; while $M^*$, $W2-W3$, and sSFR capture the star formation (see \S\ref{appendix:data} for additional information regarding the data).

\section{Causal discovery methods}
\label{sec:cau_methods}

To represent the causal structure of the dataset, we use Directed Acyclic Graphs (DAGs).
Each DAG encodes a set of conditional independencies, and DAGs that encode the same conditional independencies belong to the same Markov Equivalence Class (MEC).
This choice assumes that no cyclical dependencies between variables exist.
This is a reasonable assumption, given the clear differences in gas fractions and merger histories between the different morphological classes.
To achieve a purely data-driven study, we adopt a uniform prior, giving equal prior probability, $P(G)$, to every one of the nearly $1.14\times10^9$ possible DAGs \cite{OEIS}.
We calculate the exact posterior probabilities of every DAG given the data, $P(G \mid D)$, using the Bayesian Gaussian equivalent (BGe) score \cite{Geiger:1994,Geiger:2002,Kuipers:2014}.
The BGe score gives the marginal likelihood by examining conditional independencies and ensures that DAGs belonging to the same MEC are scored equally.

\section{Causal discovery results}
\label{sec:cau_results}

\begin{figure}
\centering
\begin{tabular}{c|c|c}
  & Most probable MEC & Corresponding DAGs \\ \hline
E & \raisebox{-0.5\height}{\includegraphics[width=0.195\textwidth]{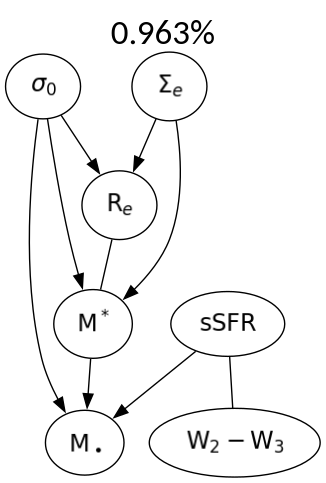}} &
\begin{tabular}{cc}
\includegraphics[width=0.18\textwidth]{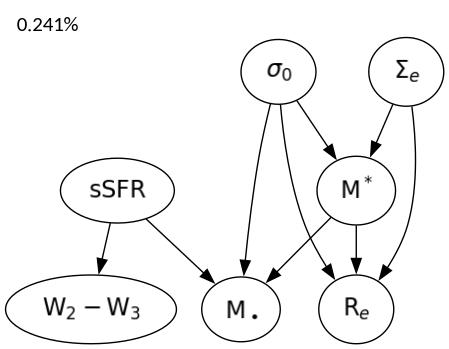} &
\includegraphics[width=0.18\textwidth]{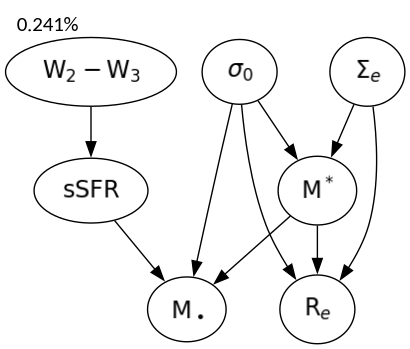} \\
\includegraphics[width=0.18\textwidth]{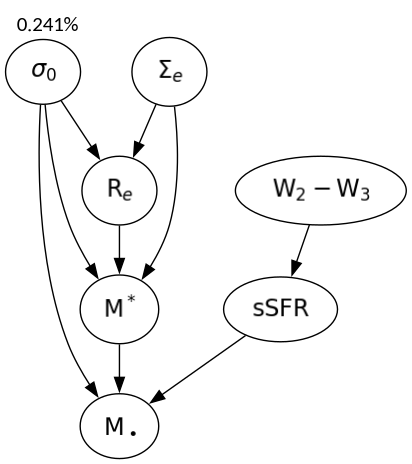} &
\includegraphics[width=0.145\textwidth]{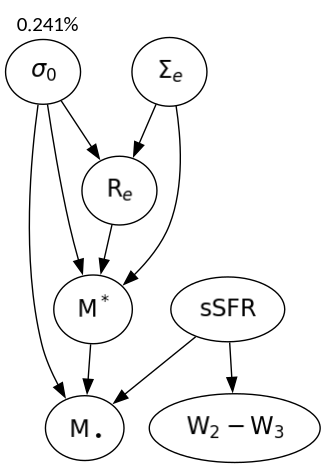} \\
\end{tabular} \\ \hline

S0 & \raisebox{-0.5\height}{\includegraphics[width=0.295\textwidth]{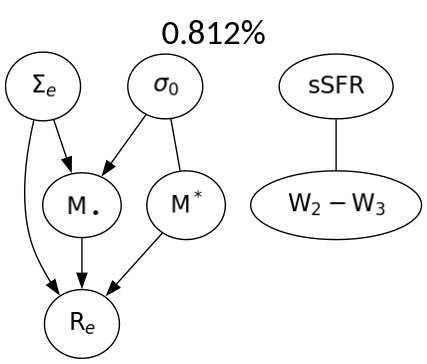}} &
\begin{tabular}{cc}
\includegraphics[width=0.195\textwidth]{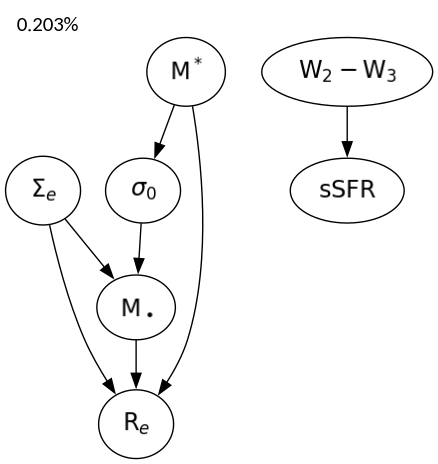} &
\includegraphics[width=0.195\textwidth]{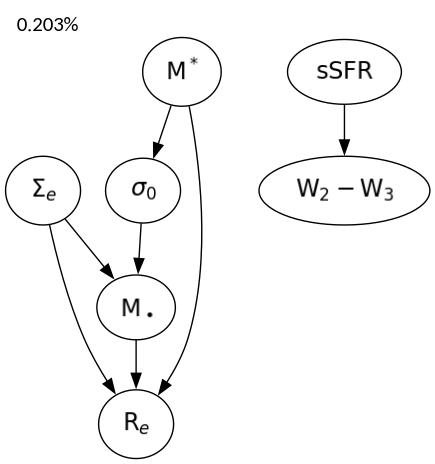} \\
\includegraphics[width=0.195\textwidth]{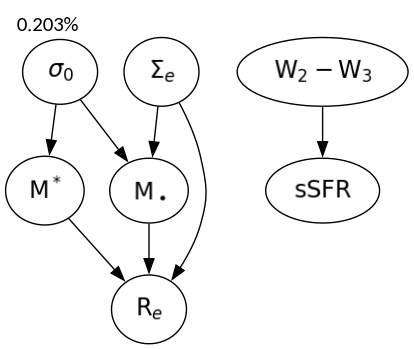} &
\includegraphics[width=0.195\textwidth]{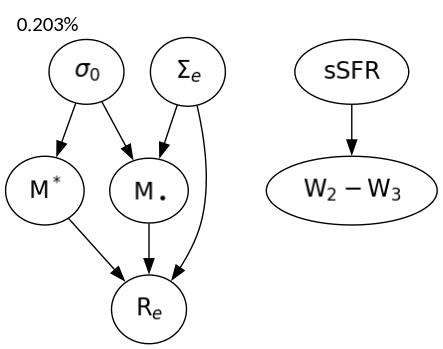} \\
\end{tabular} \\ \hline

S & \raisebox{-0.5\height}{\includegraphics[width=0.295\textwidth]{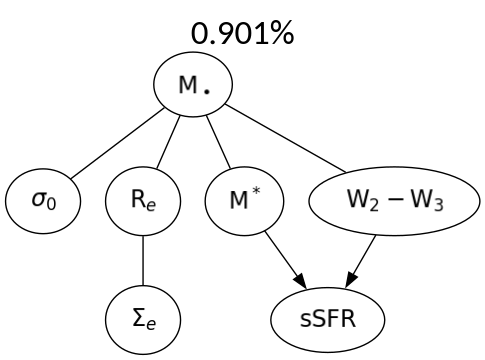}} &
\begin{tabular}{ccc}
\includegraphics[width=0.145\textwidth]{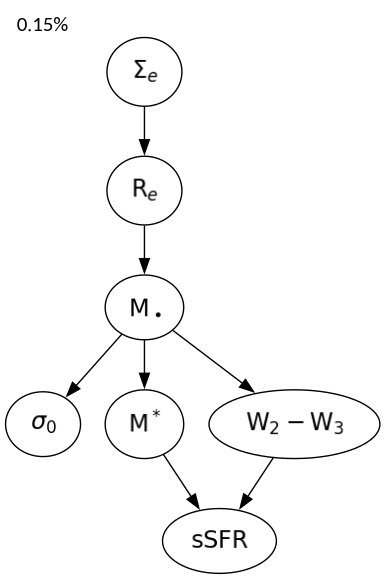} &
\includegraphics[width=0.145\textwidth]{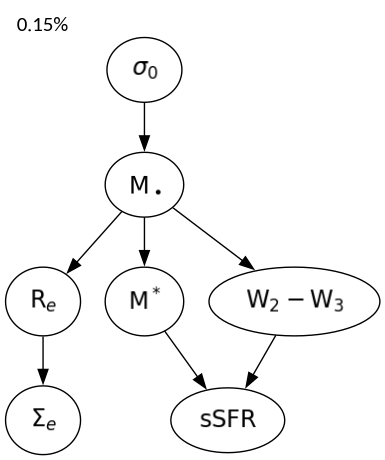} &
\includegraphics[width=0.145\textwidth]{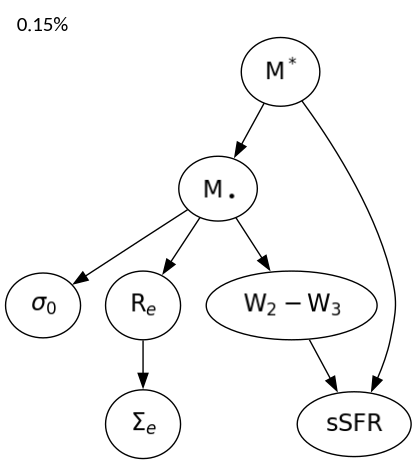} \\
\includegraphics[width=0.195\textwidth]{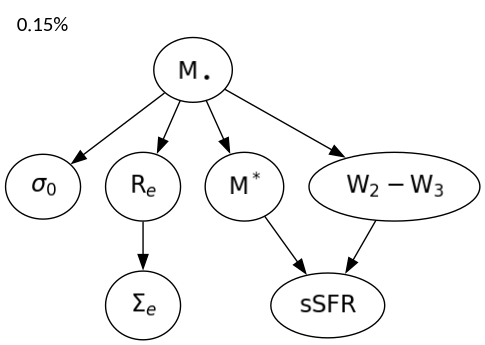} &
\includegraphics[width=0.145\textwidth]{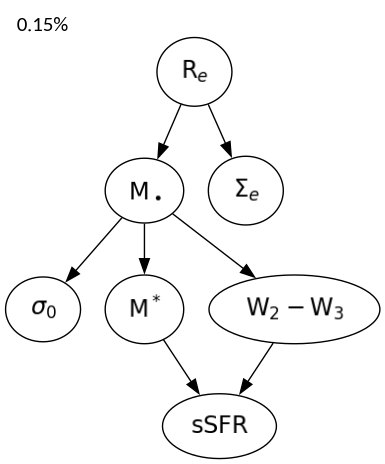} &
\includegraphics[width=0.145\textwidth]{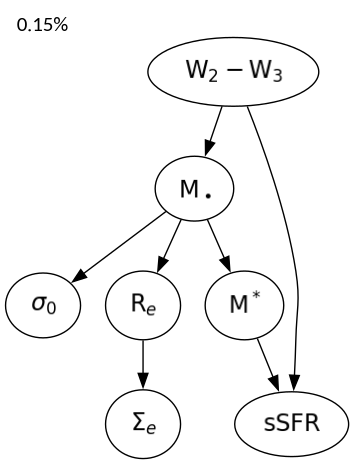} \\
\end{tabular}
\end{tabular}
\caption{
The most probable MEC for each morphology and their corresponding DAGs.
MECs are represented as Partially Directed Acyclic Graphs (PDAGs).
Directed edges suggest the direction of causality.
The undirected edge $A$ --- $B$ suggests both directions are possible (either $A\rightarrow B$ or $A\leftarrow B$), as long as no new MEC/conditional independencies are introduced by creating new colliders (i.e., two nodes both pointing towards a third node, $A\rightarrow C\leftarrow B$).
In the ellipticals, $M_\bullet$ is strictly a child, while in spiral galaxies, $M_\bullet$ is \emph{always} connected with four galaxy properties through four undirected edges, suggesting either $M_\bullet$ is the parent of all of the four galaxy properties, or $M_\bullet$ is the parent of three of the galaxy properties, and the child of the remaining one (as shown in the corresponding DAGs), ruling out more than one galaxy property pointing towards $M_\bullet$, since this creates a new collider and breaks the encoded conditional independencies.
The percentage listed above each graph indicates the posterior probability of the graph, whereas the prior probability for each individual DAG is equal to the reciprocal of the total number of DAGs (approximately $8.78\times10^{-10}$) \cite{OEIS}.
The MEC probabilities are the sum of their corresponding DAGs.
}
\label{fig:MEC+DAG}
\end{figure}

Among all possible causal structures, the most probable MEC and its corresponding DAGs for E, S0, and S galaxies are shown in Fig.~\ref{fig:MEC+DAG}.
We find that in the most probable MEC for elliptical galaxies, the SMBH mass is a causal child, i.e., an effect of galaxy properties, while in the most probable MEC for spirals, the SMBH mass is a parent of galaxy properties (with lenticulars being in the middle).

\begin{figure*}[]
  \centering
  \includegraphics[width=\linewidth]{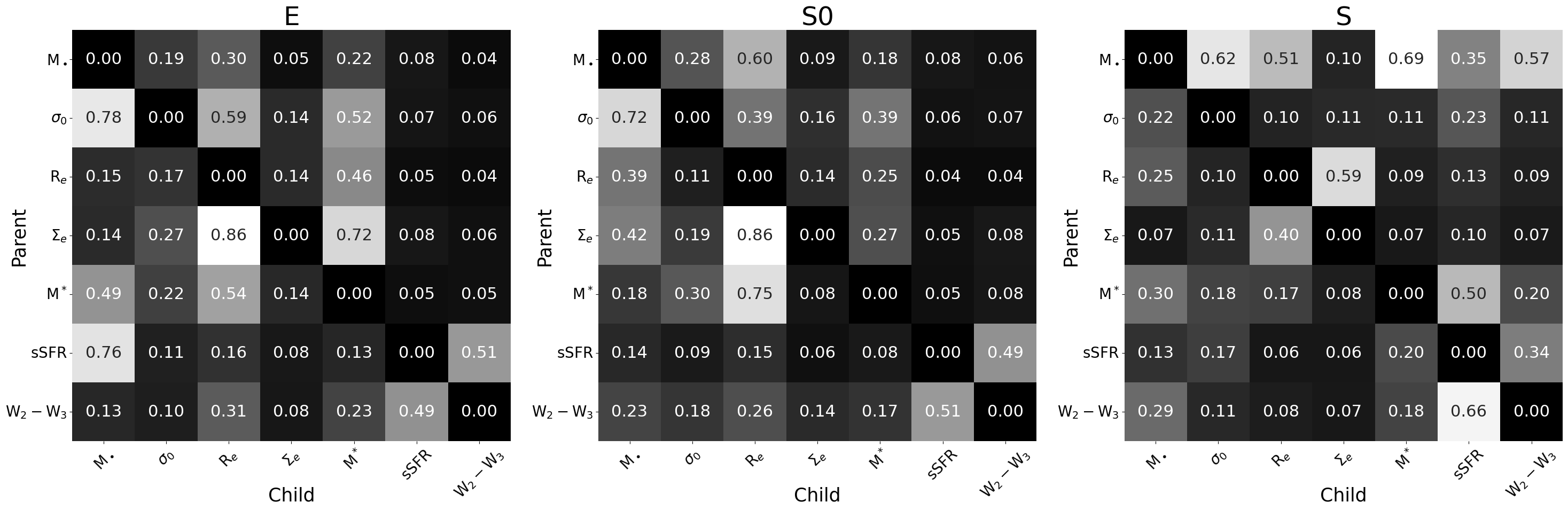}
  \includegraphics[width=\linewidth]{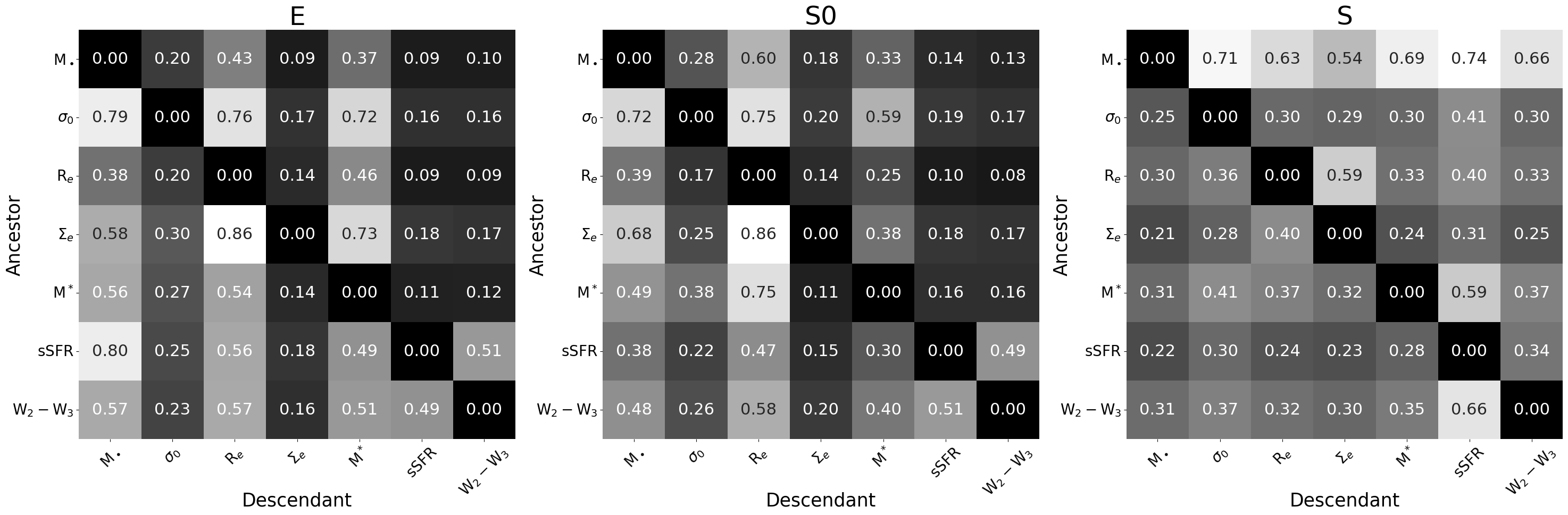}
  \caption{
  Exact posterior edge marginals (top matrices) and path marginals (bottom matrices) for elliptical (left matrices), lenticular (middle matrices), and spiral (right matrices) galaxies.
  Edge marginals give the probability of Parent $\rightarrow$ Child through directed edges summed over all DAGs and their probabilities, and path marginals give the probability of Ancestor $\rightarrow$ Descendant through both direct and indirect paths.  
  }
\label{fig:exact_matrix}
\end{figure*}

The morphologically-dependent set trend holds not only in the most probable graphs, but is common over the entire posterior distribution.
This can be quantified using edge and path marginals.
Edge marginals are the posterior probability of a direct causal relation between two variables, marginalized over the causal structures of the other nodes.
Similarly, path marginals provide the probability of a causal connection between two variables through a potentially indirect path (e.g., through intermediate nodes).
These marginal causal structures can be represented in matrix form as shown in Fig.~\ref{fig:exact_matrix}.
The first row ($M_\bullet\rightarrow$ galaxy) and column (galaxy $\rightarrow M_\bullet$) of each matrix contain information pertaining to the inferred causal relationship between SMBH masses and their host galaxy properties.

Among all possible DAGs, the percentage of graphs exhibiting a direct edge from $\sigma_0$ to $M_\bullet$ is $78\%$ in ellipticals, $72\%$ in lenticulars, and only $22\%$ in spirals.
The path marginals in the bottom row of matrices support a similar picture, as by considering all possible paths relating these two nodes, we find that $79\%$ of DAGs in ellipticals and $72\%$ in lenticulars have $\sigma_0$ as an ancestor of $M_\bullet$, whereas this is the case in only $25\%$ of DAGs in spirals.\footnote{For comparison, the null results (i.e., the posterior from a uniform prior without any data) for the edge marginals are $P(\mathrm{Parent})=29\%$, $P(\mathrm{Child})=29\%$, and $P(\mathrm{Disconnected})=42\%$; for the path marginals these probabilities are $P(\mathrm{Ancestor})=42\%$, $P(\mathrm{Descendant})=42\%$, and $P(\mathrm{Disconnected})=16\%$.}
A detailed physical interpretation of the causal structures found is presented in \S\ref{appendix:discussion}, along with discussions on unobserved confounders in \S\ref{appendix:confounders} and cyclicity in \S\ref{appendix:cyclicity}.

\section{Independent component analysis}

Astronomers characterize galaxies through properties that are selected according to subject-matter knowledge and, partly, convention.
While these can be taken for granted as a starting point for causal discovery, one of the goals of our work is also to identify potentially better representations for the dataset, in terms of coordinates that may have a clearer causal connection with the outcome of interest, i.e., the mass of the SMBH hosted by any given galaxy.
This is a causal representation learning task that in general would rely on detailed causal assumptions.
As a first step, we consider ICA as a fast and effective causal representation learning technique under the assumption of an empty causal graph, given ICA's origins as a statistical technique to transform a set of observed signals into a set of statistically independent non-Gaussian components.
This allows us to better cross-check and understand the causal structure we found in \S\ref{sec:cau_results}.
In fact, ICA has been used to infer causal structures in linear non-Gaussian settings (e.g., ICA-LiNGAM \citep{shimizu_2006}). 
In our case where the complex real world data is neither a perfect Gaussian or non-Gaussian, we exercise ICA with extra caution, keeping only non-Gaussian ICs.

\begin{figure*}[]
  \centering
  \includegraphics[width=\linewidth]{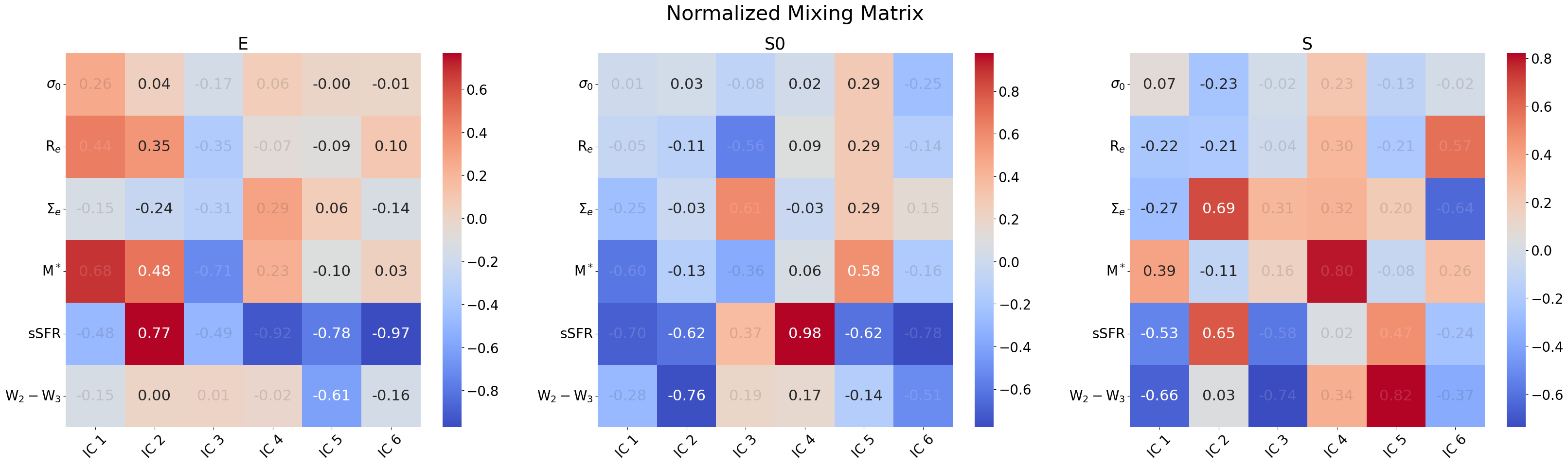}
  \includegraphics[width=\linewidth]{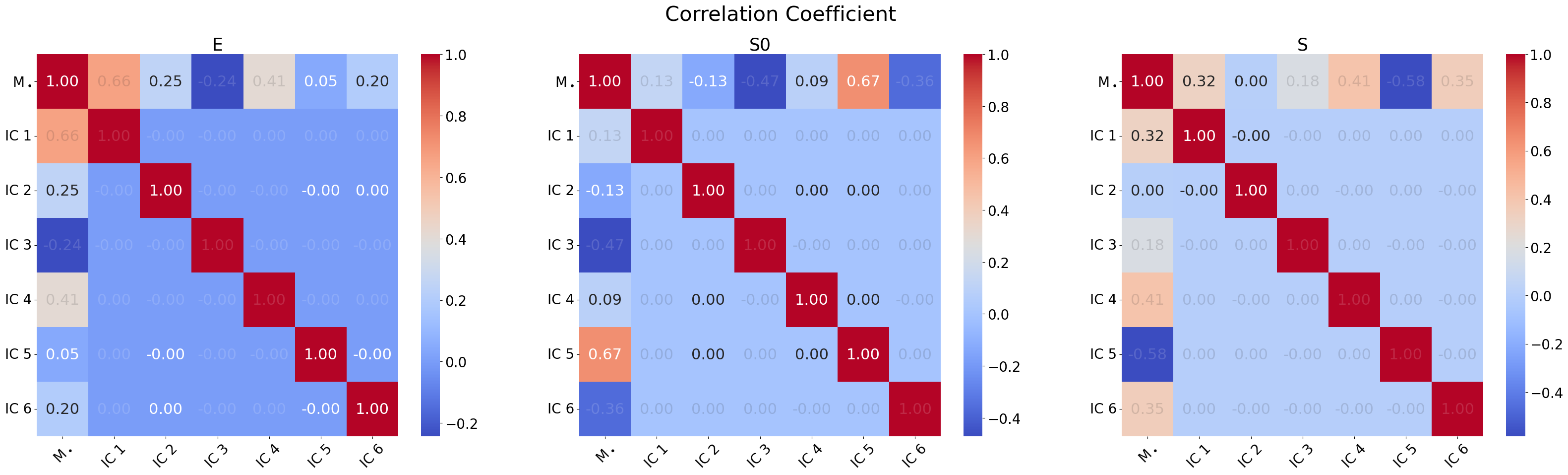}
  \includegraphics[width=\linewidth]{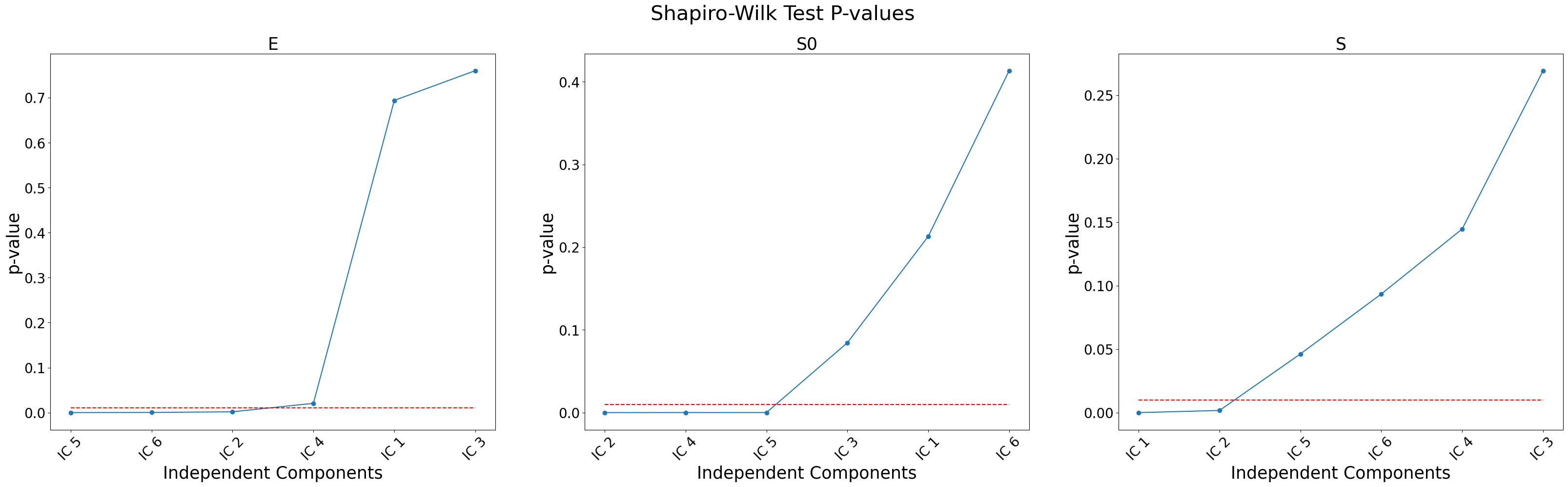}
  \caption{
  ICA, separately for E (left column), S0 (middle column), and S (right column) morphologies.
  \underline{Top row}: Unit-norm normalized mixing matrix heat map for ICs with every variable except black hole mass. 
  \underline{Middle row}: correlation coefficients between every IC and black hole mass.
  Throughout the matrices in the top two rows, we display all of the cells with meaningless ICs with faded numbers.
  \underline{Bottom row}: Shapiro-Wilk test of normality \cite{Shapiro:1965}, showing the $p$-values of every IC.
  We adopt ICs with $p$-values lower than $\alpha=0.01$ as meaningful ICs.
  }
\label{fig:ica}
\end{figure*}

We use the FastICA algorithm \citep{HYVARINEN2000411} implemented in \texttt{scikit-learn} to perform ICA on the space of six galaxy properties ($\sigma_0$, $R_\mathrm{e}$, $\Sigma_\mathrm{e}$, $M^*$, $W2-W3$, sSFR, without $M_\bullet$), with tolerance 0.0001 and function \texttt{logcosh}.
As a pre-processing step, these are all log-transformed to ensure that differences in units of measurement do not affect our results.
We run FastICA so that it outputs six ICs, but we are aware that not all of these are actually meaningful: in particular some end up having an approximately Gaussian distribution, as discussed below.
To retain only meaningful ICs, we select non-Gaussian ones by means of a Shapiro-Wilk test \cite{Shapiro:1965}, obtaining a $p$-value for each IC.
We thus retain non-Gaussian ICs, defined for our purposes as having a $p$-value $<0.01$, while we ignore the remaining ICs.
We then calculate the correlation coefficient between $M_\bullet$ and the ICs.
The normalized mixing matrix from ICA, Shapiro-Wilk test $p$-values, and correlation coefficients are shown in Fig.~\ref{fig:ica}, with Gaussian ICs grayed out.

The causal discovery method outlined in \S\ref{sec:cau_methods} \& \S\ref{sec:cau_results} reveals that galaxy properties cause $M_\bullet$ in E galaxies, while $M_\bullet$ causes galaxy properties in S galaxies.
Interestingly, ICA results confirm this finding.
Given that the ICs are independent by construction, they cannot have a common cause.
Therefore, if $M_\bullet$ causes galaxy properties, it should be statistically associated (i.e., not independent) with one IC at most.

Vice versa, if it is galaxy properties that cause $M_\bullet$, then it is possible for multiple ICs to be associated with $M_\bullet$.
These causal structures we discovered at the previous stage thus indicate that one may find multiple ICs (of galaxy properties) correlated with $M_\bullet$ in E galaxies, while only one IC correlated with $M_\bullet$ in S galaxies.
Fig.~\ref{fig:ica} shows that indeed in E galaxies, at least 2 ICs (IC$_2$ \& IC$_6$) are correlated with $M_\bullet$, and in S galaxies only IC$_1$ is correlated with $M_\bullet$, as expected.
This consistency further confirms the causal structures identified in this work.

It is worth noting that there are more meaningful ICs in E galaxies (three ICs, namely IC$_5$, IC$_6$, and IC$_2$)\footnote{Note that if instead of a fixed cutoff in $p$-value we used a visual approach such as the elbow method, the number of ICs could be raised to four, including IC$_4$ which also has a low $p$-value.} than in S galaxies (2 ICs, namely IC$_1$ and IC$_2$). 
This trend suggests that E galaxies are intrinsically more \emph{complicated} than S galaxies because they require more ICs to explain their structures.
S galaxies are clearly defined, \emph{rotationally-supported} systems that can be easily described by their more readily-measurable components related to their bulge plus large-scale disk structures.
Whereas, E galaxies are \emph{dispersion-supported} systems that consist only of a nearly featureless spheroidal component.
In this sense, E galaxies are both more \emph{complicated} (i.e., more meaningful ICs) and \emph{complex} (i.e., chaotic and disordered systems) than S galaxies.
This naturally follows the direction of increasing entropy as galaxies evolve from a more settled system (grand-design spirals) to a disorganized system (elliptical blobs of stars).

\begin{figure*}[]
  \centering
  \includegraphics[width=\linewidth]{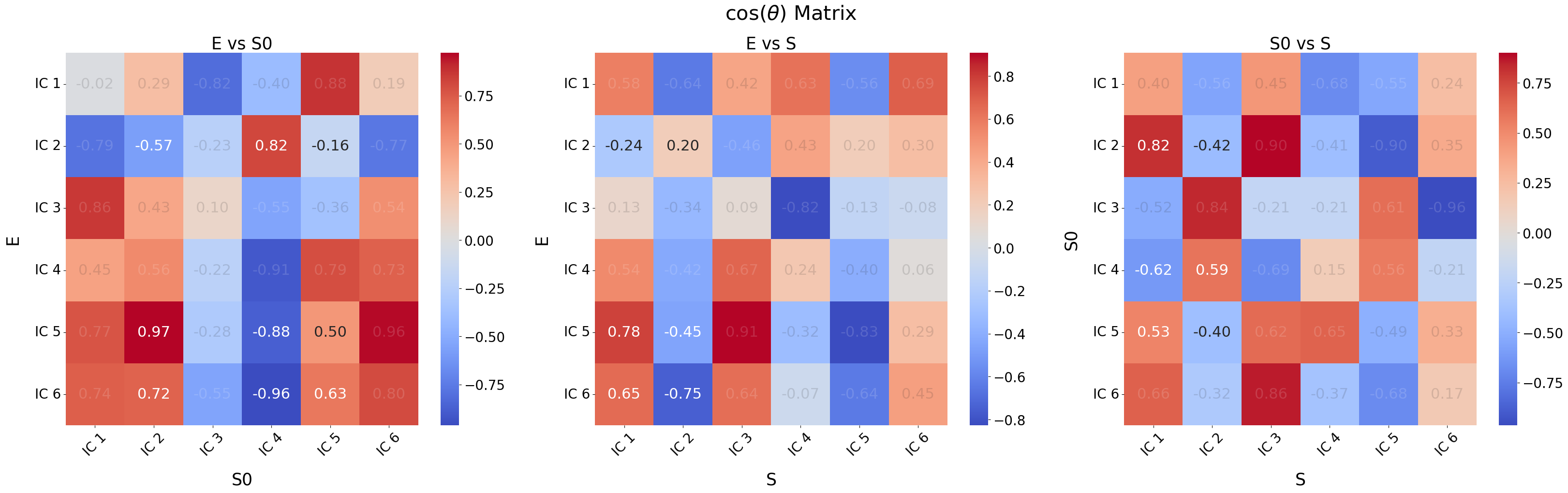}
  \caption{
  The cosine of angles between columns of mixing matrices among different morphologies.
  The cosine of the angle between two columns $u$ and $v$ is defined by $\cos\theta=\frac{u\cdot v}{|u|\cdot|v|}$.
  ICs across different morphologies are likely to carry similar interpretations when $\cos\theta$ is close to either 0 or 1.
  Similar to Fig.~\ref{fig:ica}, we gray out cells without meaningful ICs.
  }
\label{fig:ica_angle}
\end{figure*}

Interestingly, the ICs of elliptical and lenticular galaxies exhibit some clear similarities, while spiral galaxies behave entirely different.
This is shown in Fig.~\ref{fig:ica_angle} via the cosine similarity among ICs across morphologies.
IC$_5$ of elliptical galaxies essentially coincides with IC$_2$ of lenticulars, whereas IC$_6$ of ellipticals corresponds to $-$IC$_4$ of lenticulars.
Notably, these are the most meaningful ICs (the least Gaussian) both for ellipticals and for lenticulars, and are related to color and star formation rate.
No such correspondence emerges for spirals, where the physics of star formation is different due to the presence of large quantities of gas, onto which SMBH feedback impinges.

\section{Conclusions}

We have presented the first development toward causal representation learning in the form of applying ICA to an astronomical issue involving the causal interpretation of empirically observed correlations, settling a long-standing debate about whether SMBHs determine the properties of their host galaxies or vice versa.
Our main finding is that the causal direction runs from the black hole to the galaxy in the case of spiral galaxies, and in the opposite direction in the case of elliptical galaxies.
This is consistent with the established theoretical understanding that SMBHs affect galaxy evolution in the gas-rich environment of spiral galaxies through feedback (e.g., by heating the gas through jets), while galaxy properties determine the growth of SMBH mass in ellipticals due to galaxy--galaxy merger rates.

Our finding relies on applying causal discovery in the form of a score-based method with an exact posterior calculation \citep{JinSubmitted} and is confirmed by running ICA on the space of galaxy properties.
In particular, ICA shows that in spirals only one IC correlates with SMBH mass, while multiple ICs do in ellipticals, which corroborates the results of causal discovery since multiple ICs can have a combined effect but not a common cause, given that they are independent by construction.
Thus together, causal discovery and ICA enhance the interpretability of causal structures in complex datasets such as ours.
Causal discovery and ICA can synergize effectively in data analysis, particularly in disentangling multivariate datasets.
Here, we have utilized this synergy to provide a more robust interpretation of our first-of-its-kind causal discovery for astronomy by disentangling the causal structure of SMBH--galaxy coevolution.

\begin{ack}
This research was carried out on the high-performance computing resources at New York University Abu Dhabi.
We acknowledge the usage of the HyperLeda database (\url{http://leda.univ-lyon1.fr}). 
Z.J.\ and M.P.\ wish to extend their heartfelt thanks to Jithendaraa Subramanian for providing in-depth support and clarifications regarding \texttt{DAG-GFN}, and to Michelle Liu for comments and discussion.
Y.H. thanks Andrew Benson and Dhanya Sridhar for helpful discussions.
Z.J.\ thanks Michael Blanton and Joseph Gelfand for useful suggestions.
Z.J.\ genuinely thanks Mohamad Ali-Dib for his very timely help with HPC technical issues.
\textbf{Funding:}
This material is based upon work supported by Tamkeen under the NYU Abu Dhabi Research Institute grant CASS.
This work is partially supported by Schmidt Futures, a philanthropic initiative founded by Eric and Wendy Schmidt as part of the Virtual Institute for Astrophysics (VIA).
\end{ack}

\clearpage
\bibliography{scibib}
\bibliographystyle{plain}

\appendix

\section{Data}
\label{appendix:data}

\begin{figure*}[h]
  \centering
  \includegraphics[width=\linewidth]{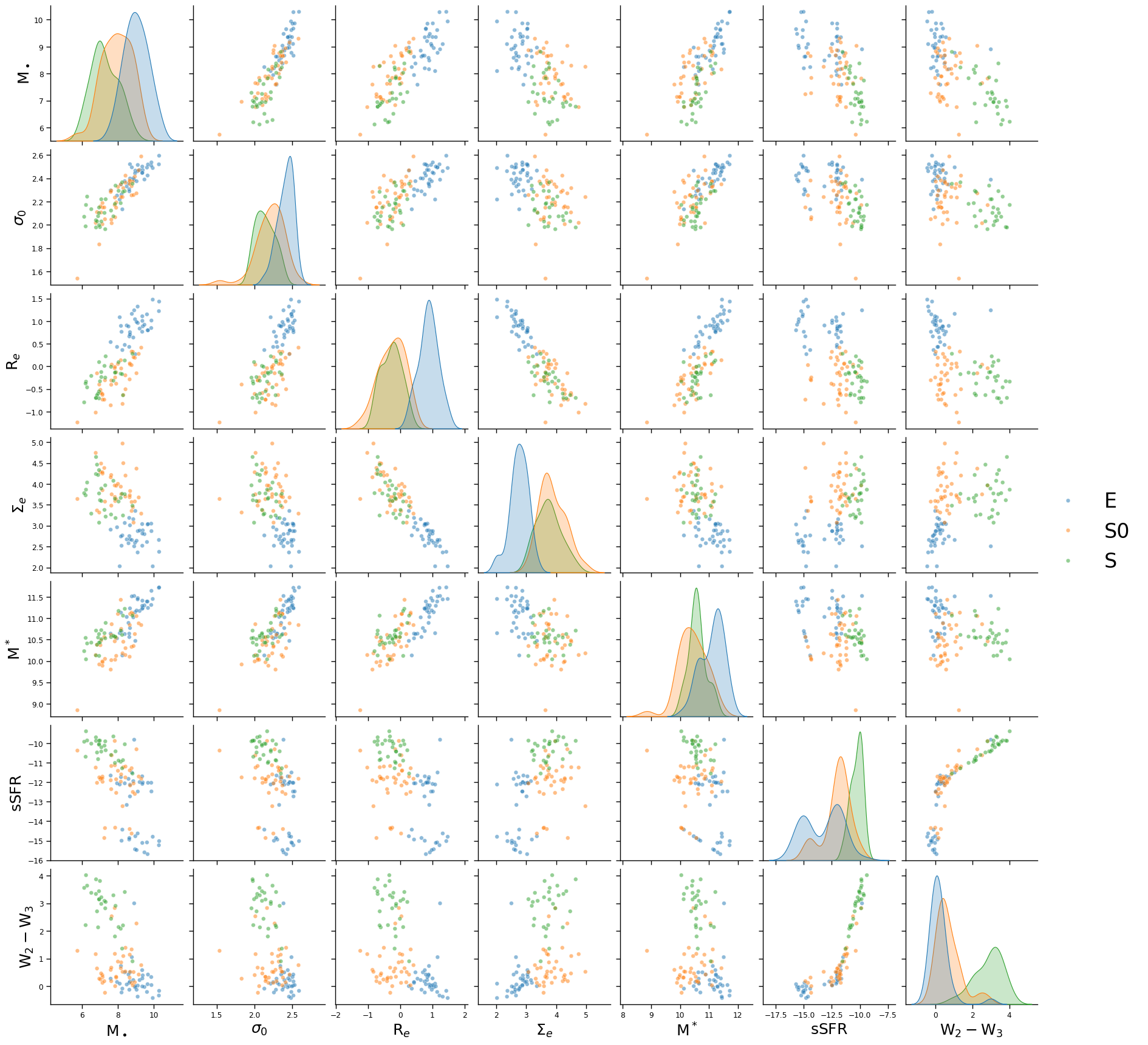}
  \caption{
  A pairplot of the data for 101 galaxies, separated into elliptical (E), lenticular (S0), and spiral (S) morphologies.
  }
\label{fig:pairplot}
\end{figure*}

$M_\bullet$ values are curated from the literature on dynamical black hole mass measurements, and $\sigma_0$ values are obtained from the HyperLeda database \cite{Makarov:2014}.
$R_\mathrm{e}$ and $\langle{\Sigma_\mathrm{e}}\rangle$ measurements come from multi-component decompositions of surface brightness light profiles (primarily of $3.6\,\mu\mathrm{m}$ \textit{Spitzer} Space Telescope imaging) from succeeding works \cite{Savorgnan:2016a,Davis:2019,Sahu:2019,Graham:2023c}.
$M^*$, $W2-W3$, and sSFR are from the Wide-field Infrared Survey Explorer, WISE \cite{Wright:2010,Graham:2024}.
The data has been used in a series of works related to black hole mass scaling relations \cite{Graham:2013,Scott:2013,Savorgnan:2013,Savorgnan:2016,Sahu:2019,Sahu:2019b,Sahu:2020,Graham:2023,Davis:2017,Davis:2018,Davis:2019,Davis:2019b,Davis:2023,Davis:2024}.
To investigate the effect of galaxy morphologies on the underlying causal structure, we further split our sample into 35 elliptical (E), 38 lenticular (S0), and 28 spiral (S) galaxies.
This choice is motivated by the observed difference in intrinsic scatter ($\epsilon$) in the $M_\bullet$--$\sigma_0$ relation \cite{Ferrarese:2000,Gebhardt:2000} in elliptical ($\epsilon=0.31$\,dex) vs.\ spiral galaxies ($\epsilon=0.67$\,dex) \cite{Sahu:2019b}, and this choice is consistent with the current understanding of quenching and hierarchical assembly \cite[e.g.,][]{Springel:2005}.
See Fig.~\ref{fig:pairplot} for a visualization of the data.

\section{Discussion on black hole and galaxy evolution}
\label{appendix:discussion}

\subsection{Causal connections for galaxy evolution}

We find that these results are consistent with theoretical models of galactic evolution.
Ellipticals are highly-evolved galaxies, being the result of a large number of galactic mergers. 
Modern hydrodynamical cosmological simulations such as IllustrisTNG \cite{Marinacci:2018,Naiman:2018,Nelson:2018,Pillepich:2018,Springel:2018} show that elliptical galaxies with $\log(M^*/\mathrm{M}_\odot)\geq11$ are generally the end result of two or more major merger events, such that the typical present-day fraction of stars with \textit{ex situ} origins is greater than 50\% \cite{Cannarozzo:2023}.\footnote{
Here, Cannarozzo et al.\ \cite{Cannarozzo:2023} follow previous work \cite{Rodriguez-Gomez:2015,Rodriguez-Gomez:2016} and define a major merger as a stellar mass ratio greater than 1/4 between the two progenitors of a given galaxy.}
In even more general terms, the process of successive mergers will act to erase the preexisting causal connection from the SMBH to its host galaxy and establish new correlations via the central limit theorem \cite{Jahnke:2011}.

During a merger, the SMBHs at the center of each merging galaxy play no role in the large-scale dynamics; it is the galaxy properties (chiefly size and mass) that shape the galaxy mergers and their outcomes.
Central SMBHs are passively driven to the bottom of the post-galaxy-merger potential well by dynamical friction, eventually merging together.
So it stands to reason to expect that in ellipticals, the distribution of SMBH masses is determined by that of galaxy properties and \emph{not} vice versa.

For spiral galaxies, this is not the case, since they experience at most a few relatively minor mergers.
Unlike elliptical galaxies, spirals are predominantly composed of \textit{in situ} stellar populations.
Causal relations between SMBH mass and galaxy properties may thus be set primordially in a secular coevolution phase, and they are not erased by mergers.
As a result, spiral galaxies behave markedly different compared to ellipticals.
Interestingly, lenticulars appear to lie in-between, as expected, based on the fact that lenticulars have undergone enough mergers to erase spiral structure while still maintaining an extended disk structure, but are not yet comparable to ellipticals in terms of mass and pressure support.\footnote{The coevolution of lenticular galaxies and their black holes is also strongly influenced by the presence of dust \cite{Graham:2023d,2024MNRAS.531..230G}.}
Moreover, by extension of Cannarozzo et al.'s results to all early-type (i.e., lenticular and elliptical) galaxies, all but the most massive lenticular galaxies should still maintain \textit{in situ} stellar fractions greater than 50\% \cite{Cannarozzo:2023}.

The six galaxy variables studied here can be split into the three parameters defining the fundamental plane (FP) of elliptical galaxies \citep{Djorgovski:1987} and three parameters related to star formation.
The FP is a manifestation of dynamical equilibrium reached in the largely pressure-supported stellar dynamics of massive elliptical galaxies \cite{Mould:2020}.
Moreover, it is a consequence of the merger formation of these galaxies via dissipation and feedback that ultimately places them on the FP.
Although only 35/101 of the galaxies are ellipticals, the classical bulges of lenticular and spiral galaxies are also governed by the FP.
Indeed, it has been found that the bulges of type S0--Sbc galaxies tightly follow the same FP relation as ellipticals \cite{Jesus:2002}.

The matrices in Fig.~\ref{fig:exact_matrix} also provide information about the causal nature of the observed FP relationship.
By looking at the path marginals for elliptical galaxies (bottom left), we find that $\langle{\Sigma_\mathrm{e}}\rangle$ is the ancestor (86\%) of $R_\mathrm{e}$ and that $\sigma_0$ is an ancestor (76\%) of $R_\mathrm{e}$.
This implies $\langle{\Sigma_\mathrm{e}}\rangle$ and $\sigma_0$ are both upstream of $R_\mathrm{e}$, confirming that the density and dynamics of stellar populations in an elliptical galaxy govern its size.
Furthermore, we find that there is nearly no chance that $M^*$ is disconnected from $R_\mathrm{e}$ (i.e., $54\%+46\%=100\%$, they are \emph{never} $d$-separated, thus \emph{always} correlated), indicating the existence of a size--mass relation due to the virial theorem (i.e., $M\sim\sigma^2R$).

\subsection{Causal active galactic nuclei feedback}

From Fig.~\ref{fig:exact_matrix}, we find that, in spirals, $M_\bullet$ is the ancestor (74\%) of sSFR, in lenticulars, there is no dominant causal direction between the two parameters (38\% and 14\%), while in ellipticals, $M_\bullet$ becomes the descendant (80\%) of the galaxy's sSFR.
This can be interpreted as a direct consequence of the presence or absence of gas through active galactic nuclei (AGN) feedback.
If there is a substantial gas reservoir (as in spirals), the SMBH is the ancestor since its feedback is responsible for shutting down star formation and hence stopping the growth of stellar mass.
With a dearth of gas, as in ellipticals, even large AGN bursts will not affect the stellar mass, and thus the SMBH cannot be an ancestor of galaxy properties.
This is further supported by the fact that we find that $M_\bullet$ is the parent (69\%) of $M^*$ in spirals, but becomes the descendant (56\%) or child (49\%) of $M^*$ in elliptical galaxies.
However, it is true that in the absence of gas, mergers are the main pathway for SMBH growth, and this will also cause the SMBH to become a descendant or child in hierarchical assembly \cite{Jahnke:2011,Graham:2023,Graham:2023b}.

We also crosscheck our results with two alternative causal discovery methods, both \emph{constraint-based}: the Peter-Clark, PC, \cite{spirtes2000causation} algorithm and the Fast Causal Inference, FCI, \cite{SpirtesManuscript-SPIAAA} algorithm, which both yield consistent results with the exact posterior approach.
Additionally, we test the inclusion of the distance to galaxies as a substitute variable, exploring the possibility of it being a hidden confounder.
We find that the causal relations identified are not altered. 
Furthermore, we find that the inferred causal relations are robust to observational errors using random sampling and to possible outlier galaxies using leave-one-out cross-validation.

The exact posterior methodology employed here for causal discovery is a powerful tool for ascertaining causal structures in a purely data-driven manner.
However, for problems with more variables, this exact approach becomes computationally intractable due to the combinatorial increase in the number of possible DAGs.
In these cases, it remains possible to quantify the posterior over DAGs through posterior samples generated with samplers such as \texttt{DAG-GFN} \cite{deleu2022daggflownet}, built on GFlowNets \cite{bengio2021gflownet,bengio2023gflownetfoundations}.
We sampled the posterior by training a \texttt{DAG-GFN}, giving results consistent with the exact-posterior approach.

Further insights can be gained by using time-series data and control variables in galaxy simulations \cite[e.g.,][]{Waterval:2024} to test the causal findings and explanations presented here. 
With knowledge of the underlying causal structures and mechanisms behind galaxy--SMBH coevolution, it should ultimately be possible to create physically-motivated black hole mass scaling relations.

We present the first data-driven evidence on the direction of the causal relationship between SMBHs and their host galaxies.
Our findings reveal that in elliptical galaxies, bulge properties influence SMBH growth, whereas in spiral galaxies, SMBHs shape galaxy characteristics.
The process of quenching can be causally explained as follows:
\begin{enumerate}
    \item quenching starts in gas rich (i.e., spiral) galaxies, and hence there is a causal connection
    \item the quenching is over in elliptical galaxies, where we only see the end product of such quenching, and the causal connection is now reversed.
\end{enumerate} 
These findings support theoretical models of galactic evolution driven by feedback processes and mergers.
The successful application of causal discovery to this astrophysical dataset paves the way for a deeper understanding of the fundamental physical processes driving galaxy evolution and establishes causal discovery as a powerful tool for data-driven breakthroughs across various scientific disciplines.

\section{Discussion on unobserved confounders}\label{appendix:confounders}

Our posterior calculation approach implicitly adopts the assumption of causal sufficiency, i.e., assuming there are no unobserved confounders\footnote{An unobserved confounder is a variable that is not included in the analysis, but is a cause of two or more variables of interest.}. 
With the presence of an unobserved confounder, non-existing causal relations might be falsely identified.
Some potential unobserved confounders, such as the reserve of gas or merger history, are practically difficult to observe but are already integrated into our interpretation.
However, the distance from us to galaxies does not directly play any role in galaxy formation theory nor in our interpretation, but might influence multiple variables we examined, since our ability to measure all these seven variables decreases as distance increases and thus bias our sample towards nearby and more massive BHs/galaxies.
Therefore, we examined the impact of distance by performing causal discovery with distance as one of the seven variables.

\begin{figure*}
  \centering
  \includegraphics[width=\linewidth]{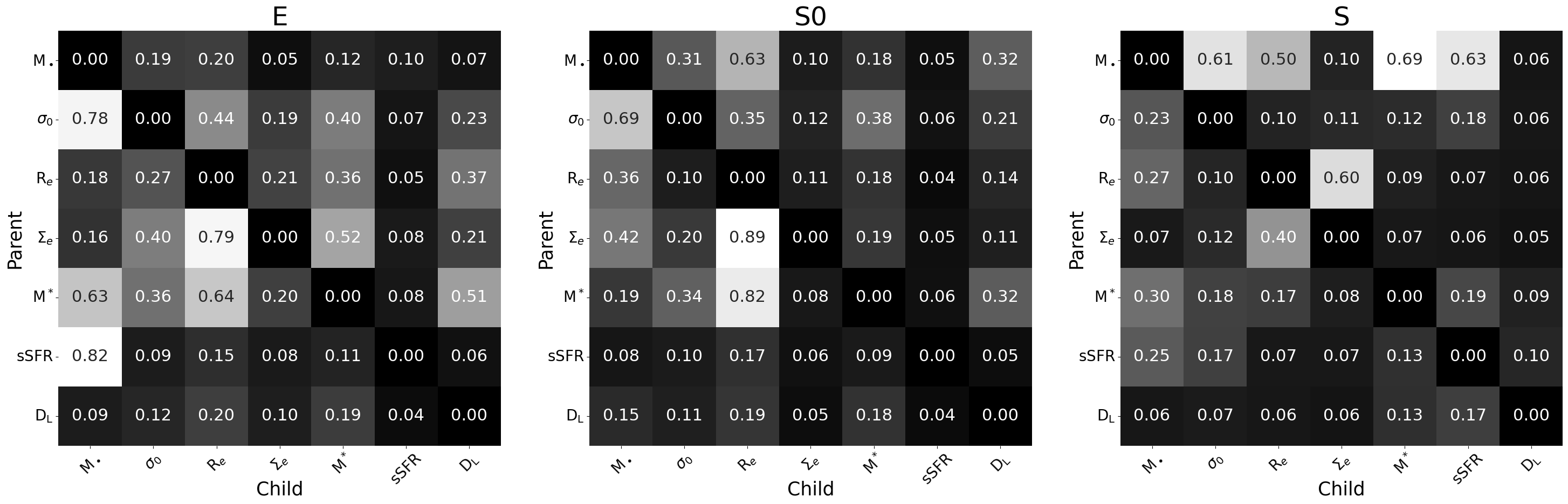}
  \includegraphics[width=\linewidth]{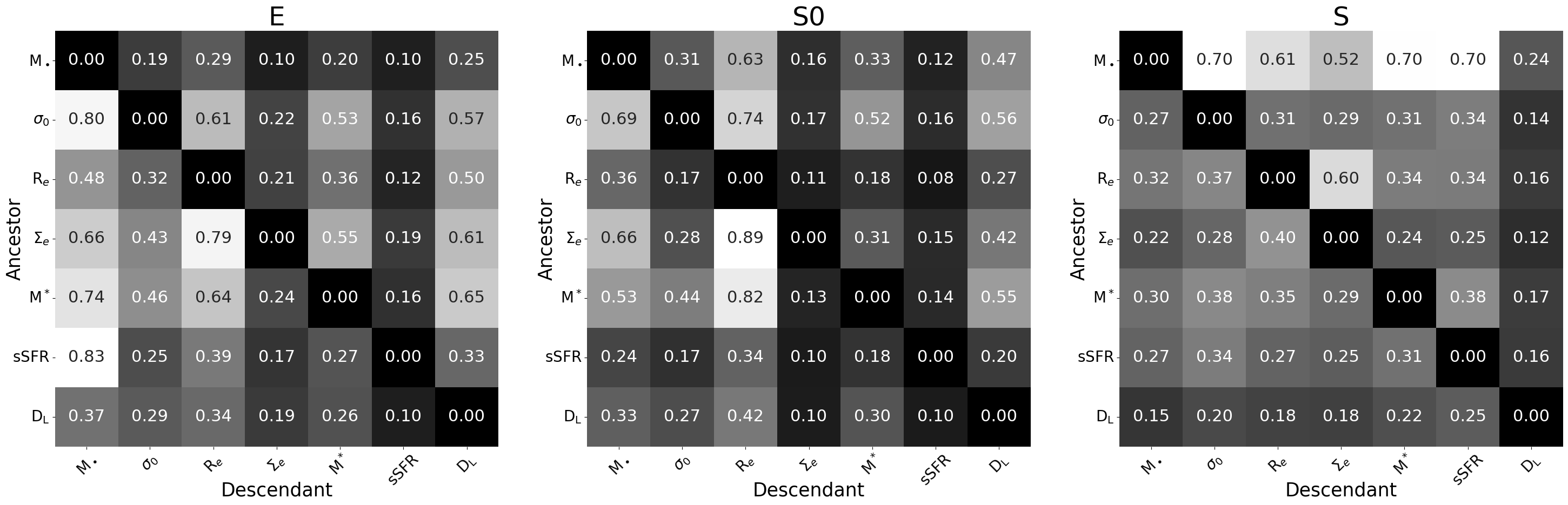}
  \caption{
  Edge marginals (top matrices) and path marginals (bottom matrices) with luminosity distance ($D_L$) as one of the variables.
  }
\label{fig:distance}
\end{figure*}

Since $W2-W3$ and sSFR are highly degenerate with each other, we replaced $W2-W3$ with $D_L$, the luminosity distance to the targets\footnote{
The luminosity distances are adopted from \cite{Graham:2023}.
Indeed, this sample of dynamically-measured black hole masses comes from galaxies that are all in the local Universe (median $D_L=19.3$\,Mpc; $z=0.00439$ according to \cite{Planck:2020}).}.
The edge and path marginals with distance included are presented in Fig.~\ref{fig:distance}.
Comparing against the original marginals without distance (Fig.~\ref{fig:exact_matrix}), the presence of distance barely changes any previously identified causal relations, since the edge and path marginals between galaxy properties and SMBH masses remains unchanged with or without the inclusion of distance.

\section{Discussion on cyclicity}\label{appendix:cyclicity}

By calculating the posterior probabilities of all possible DAGs, we implicitly assumed acyclicity, i.e., no loops in a graph.
In fact, the existence of feedback loops between black hole mass and galaxy properties (i.e., having black hole mass causing the galaxy properties, and then galaxy properties also causing black hole mass at the same time) is trivial in ellipticals and spirals according to galaxy formation theory. 
Black holes affect their host galaxies through black hole feedback, a process that heats the gas and pushes gas out to starve star formation, while galaxies also affect the central black hole through mergers and accretion.
In an ideal spiral galaxy, there have been (at most) only minor mergers, thus killing off the merger path of galaxy $\rightarrow$ black hole.

\begin{figure*}
  \centering
\includegraphics[width=\linewidth]{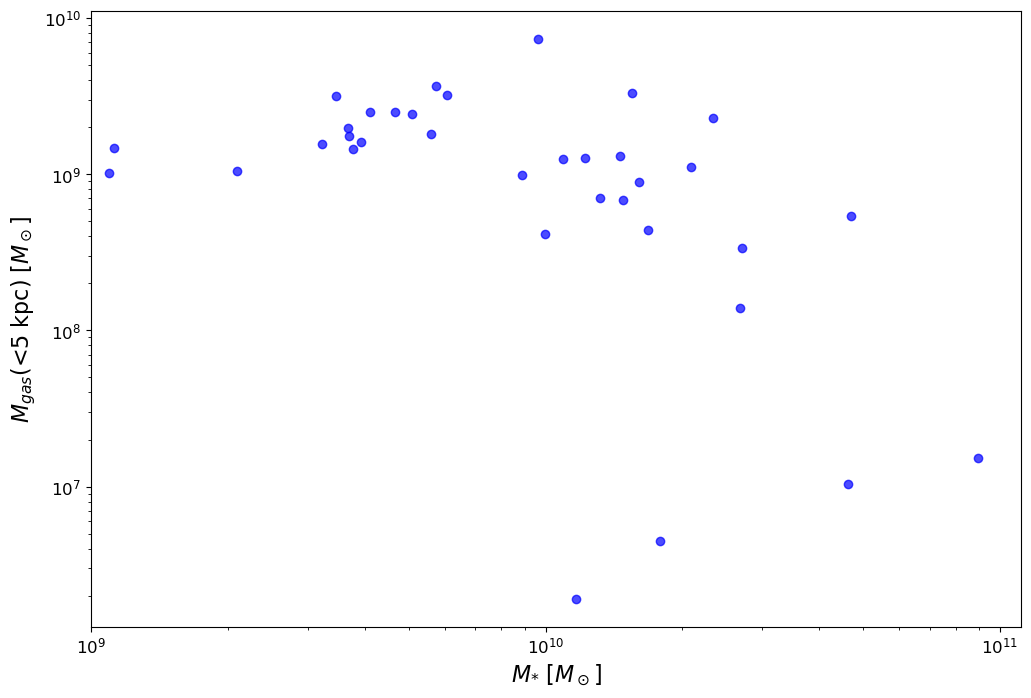}
  \caption{
  Gas mass within 5\,kpc versus total stellar mass in NIHAO simulated galaxies \citep{Wang2015}.
  The central gas mass is fairly constant in gas-rich galaxies, implying that gas accretion onto the black hole, which is mainly regulated by the black hole mass and the local gas density \citep{Bondi1952}, is also quite uniform across galaxies, weakening the galaxy $\rightarrow$ SMBH causal relation in spirals.
  }
\label{fig:NIHAO}
\end{figure*}

The accretion onto the black hole is mainly regulated by the black hole mass itself and the gas density in the central region \citep{Bondi1952}.
This latter quantity is found to be relatively constant in gas-rich galaxies, as confirmed by modern numerical simulations, like the NIHAO suite \citep{Wang2015,Blank2019} as shown in Figure~\ref{fig:NIHAO}.
This implies that accretion is fairly constant in all gas-rich galaxies, diminishing the causal relation galaxy $\rightarrow$ black hole.

Therefore, in spiral galaxies, the causal relation of galaxy $\rightarrow$ black hole is expected to be very weak compared to the black hole $\rightarrow$ galaxy direction.
On the other hand, ellipticals are in short supply of gas, therefore the central SMBH lacks the media in which to project its energy to regulate star formation.
As a result, in ellipticals, the black hole $\rightarrow$ galaxy direction is negligible compared to the galaxy $\rightarrow$ black hole path enabled by major mergers.

In all (in spirals and ellipticals), one of the causal directions between SMBHs and galaxies is expected to considerably overwhelm the other, making the causal structure acyclic.
The lenticulars, however, might have both major mergers and black hole feedback simultaneously, thus being more cyclic in their causal structure.
To fully identify cyclic causal structures, time-series data is usually required.
While in our case of SMBH--galaxy coevolution, which happens on a timescale of billions of years, obtaining time-series data is impossible within the lifetime of humanity\footnote{Except in simulations, which we will investigate in future work.}, studies of samples of galaxies with different ages may provide observational clues about the presence or absence of cyclicity in future studies.

\end{document}